\documentclass[10pt,preprint2]{aastex}

\usepackage{bm}
\usepackage{textcomp}
\usepackage{graphicx}
\usepackage{amssymb}
\usepackage{gensymb}

\begin{document}

\date{}
\title{Polarized light scattering with Paschen-Back effect, level-crossing of fine
structure states and partial frequency redistribution}
\author{K. Sowmya$^{1}$, K. N. Nagendra$^{1}$, M. Sampoorna$^{1}$ and
J. O. Stenflo$^{2,3}$} 
\affil{$^1$Indian Institute of Astrophysics, Koramangala, Bengaluru, India}
\affil{$^2$Institute of Astronomy,
ETH Zurich, CH-8093 Zurich, Switzerland }
\affil{$^3$Istituto Ricerche Solari Locarno, Via Patocchi,
6605 Locarno-Monti, Switzerland}

\email{ksowmya@iiap.res.in; knn@iiap.res.in; sampoorna@iiap.res.in;
stenflo@astro.phys.ethz.ch}
\date{}

\begin{abstract}
The quantum interference between the fine structure states of an atom modifies the
shapes of the emergent Stokes profiles in the Second Solar Spectrum. This phenomenon
has been studied in great detail both in the presence and absence of magnetic fields.
By assuming a flat-spectrum for the incident radiation, the signatures of this effect
have been explored for arbitrary field strengths. Even though the theory which takes
into account the frequency dependence of the incident radiation is well developed, it
is restricted to the regime in which the
magnetic splitting is much smaller than the fine
structure splitting. In the present paper, we carry out a generalization of our
scattering matrix formalism including the effects of partial frequency redistribution
(PRD) for arbitrary magnetic fields. We test the formalism using available benchmarks
for special cases. In particular we apply it to the Li\,{\sc i} 6708\,\AA\ D$_1$ and
D$_2$ line system, for which observable effects from the Paschen-Back regime are
expected in the Sun's spectrum.
\end{abstract}


\section{Introduction}
\label{intro}
The interaction of the spin of the electrons with their orbital angular
momenta leads to splitting of the atomic levels into fine structure states
that are labeled by the total electronic angular momentum quantum numbers $J$.
The magnetic substates belonging to these $J$ states are degenerate in
the absence of a magnetic field. 
When a magnetic field is applied, the degeneracy is lifted and the energies of these
magnetic substates are modified. 
With an increase in the field strength, the magnetic substates belonging to different
$J$ states of 
a given term begin to overlap leading to a mixing
of the $J$ states and $J$ no longer remains a good quantum number. The Paschen-Back effect
(PBE) occurs when the
splitting produced by the magnetic field is comparable to the fine
structure splitting. 

In the present paper we address the problem of PBE on a two-term atom taking into account 
the effects of partial frequency redistribution (PRD). 
In other words, we study the $J$-state interference phenomenon in the
presence of a magnetic field of arbitrary strength. 
In particular we derive the PRD 
matrix for the problem at hand and present the results
obtained for the single scattering case.

\citet{bom80} developed a density matrix formalism to handle $J$-state 
interference in the presence of a magnetic field of arbitrary strength 
(including both the Zeeman and the Paschen-Back regimes). Her formalism was 
limited to the complete frequency redistribution (CRD) in scattering. 
A quantum theory of $J$-state
interference phenomenon for the case of frequency coherent
scattering was formulated by \citet{s80,s94,s97}.
Under the flat-spectrum approximation, \citet[][hereafter LL04]{ll04}
developed a QED theory for the $J$-state interference phenomenon in
a multi-term atom and in the presence of magnetic 
fields of arbitrary strengths. Assuming 
CRD, \citet{cm05} considered the problem of PBE in a multi-term atom involving 
the interferences among both $J$ and $F$ states. In the linear Zeeman regime, where the
fine structure splitting is larger than the splitting produced by the magnetic field,
\citet{smi11a,smi13} developed a theory for interference between the fine structure
states taking into account the effects of PRD. In the present paper, we generalize the 
collisionless redistribution matrix (hereafter RM) derived by \citet{smi11a} to include
the PBE. In other words, we present
a general form of the RM which holds good in Hanle, Zeeman as well as PB regimes.

PBE is of great interest to the stellar as well as 
the solar community because it serves as an 
effective tool to diagnose vector magnetic fields. The emergent spectral line
polarization depends sensitively on the magnetic field. PBE in molecules has proven to be
a good diagnostic  tool in recent years for magnetic field measurements. Since the
fine structure splittings in molecules are smaller than those for atoms, the PBE
becomes operative for relatively lower field strengths in molecules. 
Attempts have been made to develop a theoretical framework
for this problem and to identify and understand the signatures of this effect in the
emergent line polarization \citep[see for e.g.,][]{as06,ber05,ber06,ber06a,shapetal06,
shapetal07}. The important step is to set up the Hamiltonian in the right form and
diagonalize it to get the energy eigenvalues and eigenvectors which can be used later
in the computations of the Stokes parameters. To this end, in our work concerned with
PBE in atoms, we use the diagonalization code of \citet{landi78}. This is a computer
program written for the simultaneous diagonalization of the magnetic and the
hyperfine structure Hamiltonian. We modify this program suitably for the problem at hand.

In the work on PBE on hyperfine structure states of a two-level atom \citep[see][]{sow14}, 
we derived the PRD matrix for this process. Further, we studied the characteristics of the
RM in a single 90\textdegree\ scattering event. The same
framework can also be developed for the case of PBE on fine structure states with
the straightforward replacement of the quantum numbers which we discuss in
Section~\ref{pbl}.
In Section~\ref{atom-sys} we set up the total Hamiltonian for PBE in a two-term atom.
The Hamiltonian in this case has non-zero non-diagonal elements which represent the
mixing of the $J$ states. The general form of the RM in terms of the irreducible
spherical tensors, derived assuming the lower levels to be unpolarized and infinitely
sharp, is presented in Section~\ref{pbl}. The results for the single scattering case
are discussed in Section~\ref{results} considering the fine structure states of 
the lithium D$_1$ and D$_2$ lines as an example. 
In the solar case, the Li\,{\sc i} 6708\,\AA\ doublet, 
which has the same fine structure configuration as the D$_1$ and D$_2$ lines of 
Na\,{\sc i} and Ba\,{\sc ii}, but for which the fine
structure splitting is only 0.15\,\AA, serves as a good candidate for application of the
theory developed in the present paper. Spectropolarimetric observations 
of this Li\,{\sc i} doublet have been published in \citet{s11}. The theoretical 
work on the same spectral line system has been presented by \citet{bel09} in the limit of
microturbulent fields and for the non-magnetic case. Section~\ref{conclu} is devoted for
concluding remarks.

\section{PBE in a two-term atom}
\label{atom-sys}
We consider a two-term atom described by the $L-S$ coupling scheme.
Under the $L-S$ coupling approximation, the fine structure Hamiltonian is given by
\begin{eqnarray}
 && \mathcal{H}_{\rm fs}=\zeta(L_kS) {\bm L_k}\cdot{\bm S}\ ,
\label{h-fs}
\end{eqnarray}
where $\zeta(L_kS)$ has the dimensions of energy and is given by the
`Land\'e-interval' rule as
\begin{eqnarray}
 && \zeta(L_kS)=\frac{E(J_k)-E(J_k-1)}{J_k}\ .
\label{zeta}
\end{eqnarray}
Here $k=a$ (lower term) or $b$ (upper term).
The energy shift due
to spin-orbit coupling can be obtained from the Hund's rule 3 as
\begin{eqnarray}
 && E_{L_kS}(J_k)=\frac{1}{2}\zeta(L_kS)
\nonumber \\ &&
\!\!\!\!\!\times[J_k(J_k+1)-L_k(L_k+1)-S(S+1)]\ .
\label{dele}
\end{eqnarray}
If an external magnetic field is applied then its interaction
with the atomic system is described by the Hamiltonian
\begin{eqnarray}
 && \mathcal{H}_B=\mu_0 ({\bm L_k}+2{\bm S})\cdot{\bm B}\ ,
\label{h-mag}
\end{eqnarray}
where $\mu_0$ is the Bohr-magneton. 
If the applied magnetic field produces a splitting comparable to the fine
structure splitting, then the magnetic Hamiltonian can no longer be treated as
a perturbation to the spin-orbit Hamiltonian, $\mathcal{H}_{\rm fs}$. In this case the
energy levels have to be found by diagonalizing the total Hamiltonian
$\mathcal{H}$ given by
\begin{eqnarray}
 && \mathcal{H}=\mathcal{H}_{\rm fs}+\mathcal{H}_{B}\ .
\label{htot}
\end{eqnarray}
The quantization axis (z-axis) is taken to be along the applied magnetic field so
that the total Hamiltonian can be diagonalized in the energy
eigenvector basis $|L_kSJ_k\mu_k\rangle$. However in the PB regime the magnetic field 
produces a mixing of the $J$ states belonging to a given term. Thus the eigenvectors of 
the total Hamiltonian are of the form
\begin{equation}
|L_kSj_k\mu_k\rangle = \sum_{J_k} C^{j_k}_{J_k}(L_kS,\mu_k) |L_kSJ_k\mu_k\rangle,
\label{pb-eigne-vector}
\end{equation}
where the symbol $j_k$ labels different states spanned by
the quantum numbers $(L_k,S,\mu_k)$ and\\
\noindent$C^{j_k}_{J_k}(L_kS,\mu_k)$ are the expansion
coefficients. To determine the eigenvectors 
$|L_kSj_k\mu_k\rangle$ and the corresponding eigenvalues we have to 
diagonalize a set of matrices of the form
\begin{equation}
 \langle L_kSJ_k\mu_k|\mathcal{H}_{\rm fs}+\mathcal{H}_{B}|L_kSJ_{k^\prime}\mu_k\rangle\ .
\label{mat-diag}
\end{equation}
The above expression indicates that a given $\mu_k$ can be assigned to both $J_k$
and $J_{k^\prime}$ as a result of level interference.
Since the spin orbit Hamiltonian is diagonal in $J_k$ we have
\begin{eqnarray}
 && \langle L_kSJ_k\mu_k|\mathcal{H}_{\rm fs}|L_kSJ_k\mu_k\rangle = E_{L_kS}(J_k)\ ,
\label{h1}
\end{eqnarray}
where $E_{L_kS}(J_k)$ is given by Equation~(\ref{dele}). The magnetic Hamiltonian 
can be written in the energy eigenvector basis as
\begin{eqnarray}
 && \!\!\!\!\!\!\!\!\!\!
\langle L_kSJ_k\mu_k|\mathcal{H}_{B}|L_kSJ_{k^\prime}\mu_{k^\prime}\rangle=
\delta_{\mu_k\mu_{k^\prime}}\mu_0B\nonumber \\ && 
\!\!\!\!\!\!\!\!\!\!\times\bigg[\mu_k\delta_{J_kJ_{k^\prime}}+(-1)^{J_k+J_{k^\prime}+L_k+S+\mu_k}
\nonumber \\ &&\!\!\!\!\!\!\!\!\!\!\times
\sqrt{(2J_k+1)(2J_{k^\prime}+1)S(S+1)(2S+1)}\nonumber \\ &&\!\!\!\!\!\!\!\!\!\!\times
\left (
\begin{array}{ccc}
J_k & J_{k^\prime} & 1 \\
-\mu_k & \mu_k & 0 \\
\end{array}
\right ) 
\left\lbrace 
\begin{array}{ccc}
J_k & J_{k^\prime} & 1\\
S & S & L_k \\
\end{array}
\right\rbrace\ \bigg]\ .
\label{h2}
\end{eqnarray}
The diagonalization of the total Hamiltonian gives the energy eigenvalues
and the energy eigenvectors \citep[see][]{landi78}. For simplicity we consider
the PBE only in the upper term and neglect the crossing of magnetic substates
belonging to different fine structure states in the lower term.

\section{PRD matrix for the Paschen-Back effect on fine structure states}
\label{pbl}
The steps followed in deriving the RM 
are the same as those in the case of PBE in hyperfine structure states
\citep[see][]{sow14}.
The resulting RM for $J$-state interference in the presence of magnetic fields
of arbitrary strengths can also be obtained from the corresponding RM
for the $F$-state interference phenomenon by the following
quantum number replacement:
\begin{equation}
F\rightarrow J;\quad J\rightarrow L;\quad I_s\rightarrow S;\quad i\rightarrow j\ ,
\label{substn}
\end{equation}
in the latter RM. Here $F$ ($={\bm {J+I_s}}$) is the total angular momentum,
$J$ ($={\bm {L+S}}$)
is the total electronic angular momentum, $L$ is
the orbital angular momentum, $I_s$ is the nuclear spin and $S$ is the electron spin
angular momentum. $i$ and $j$ label different states spanned by the quantum
numbers $(J,I_s,\mu_F)$ and $(L,S,\mu_J)$ respectively. Here $\mu_F$ and $\mu_J$
are the projections of $F$ and $J$ on the quantization axis.
Thus the RM for $J$-state interference in the presence of a magnetic field of arbitrary
strength can be written as
\begin{eqnarray}
&&\!\!\!\!\!\!\!\!\!\!{\bf R}_{ij}^{\rm II}
(x,{\bm n},x^\prime,{\bm n}^\prime;{\bm B})=
\frac{3(2L_b+1)}{(2S+1)}
\nonumber \\ && \!\!\!\!\!\!\!\!\!\! \times
\sum_{KK^\prime Q}
\sum_{j_a\mu_aj_f\mu_fj_b\mu_bj_{b^\prime}\mu_{b^\prime}}
\nonumber \\ && \!\!\!\!\!\!\!\!\!\! \times
\sum_{J_aJ_{a^\prime}J_fJ_{f^\prime}J_bJ_{b^\prime}
J_{b^{\prime\prime}}J_{b^{\prime\prime\prime}}}
\sum_{qq^\prime q^{\prime\prime}
q^{\prime\prime\prime}}
(-1)^{q-q^{\prime\prime\prime}+Q}
\nonumber \\ && \!\!\!\!\!\!\!\!\!\! \times
\sqrt{(2K+1)(2K^\prime+1)}
\cos\beta_{j_{b^\prime}\mu_{b^\prime}j_b\mu_b}
{\rm e}^{{\rm i}\beta_{j_{b^\prime}\mu_{b^\prime}j_b\mu_b}}
\nonumber \\ && \!\!\!\!\!\!\!\!\!\! \times
[(h^{\rm II}_{j_b\mu_b,j_{b^\prime}\mu_{b^\prime}})_{j_a\mu_aj_f\mu_f}+
{\rm i}(f^{\rm II}_{j_b\mu_b,j_{b^\prime}\mu_{b^\prime}})_{j_a\mu_aj_f\mu_f}]
\nonumber \\ && \!\!\!\!\!\!\!\!\!\! \times 
C^{j_f}_{J_f}(L_aS,\mu_f) C^{j_a}_{J_a}(L_aS,\mu_a)
C^{j_b}_{J_b}(L_bS,\mu_b) 
\nonumber \\ && \!\!\!\!\!\!\!\!\!\! \times
C^{j_b}_{J_{b^{\prime\prime}}}(L_bS,\mu_b)
C^{j_f}_{J_{f^\prime}}(L_aS,\mu_f) C^{j_a}_{J_{a^\prime}}(L_aS,\mu_a)
\nonumber \\ && \!\!\!\!\!\!\!\!\!\! \times
C^{j_{b^\prime}}_{J_{b^\prime}}(L_bS,\mu_{b^\prime})
C^{j_{b^\prime}}_{J_{b^{\prime\prime\prime}}}(L_bS,\mu_{b^\prime})
\nonumber \\ && \!\!\!\!\!\!\!\!\!\! \times
\sqrt{(2J_a+1)(2J_{a^\prime}+1)(2J_f+1)(2J_{f^\prime}+1)}
\nonumber \\ && \!\!\!\!\!\!\!\!\!\! \times
\sqrt{(2J_b+1)(2J_{b^\prime}+1)(2J_{b^{\prime\prime}}+1)
(2J_{b^{\prime\prime\prime}}+1)}
\nonumber \\ && \!\!\!\!\!\!\!\!\!\! \times
\left (
\begin{array}{ccc}
J_b & J_f & 1\\
-\mu_b & \mu_f & -q \\
\end{array}
\right )
\left (
\begin{array}{ccc}
J_{b^\prime} & J_{f^\prime} & 1\\
-\mu_{b^\prime} & \mu_f & -q^{\prime} \\
\end{array}
\right )
\nonumber \\ && \!\!\!\!\!\!\!\!\!\! \times
\left (
\begin{array}{ccc}
J_{b^{\prime\prime}} & J_a & 1\\
-\mu_b & \mu_a & -q^{\prime\prime} \\
\end{array}
\right )
\left (
\begin{array}{ccc}
J_{b^{\prime\prime\prime}} & J_{a^\prime} & 1\\
-\mu_{b^\prime} & \mu_a & -q^{\prime\prime\prime} \\
\end{array}
\right )
\nonumber \\ && \!\!\!\!\!\!\!\!\!\! \times
\left (
\begin{array}{ccc}
1 & 1 & K\\
q & -q^{\prime} & -Q \\
\end{array}
\right )
\left (
\begin{array}{ccc}
1 & 1 & K^\prime\\
q^{\prime\prime\prime} & -q^{\prime\prime} & Q\\
\end{array}
\right )
\nonumber \\ && \!\!\!\!\!\!\!\!\!\! \times
\left\lbrace
\begin{array}{ccc}
L_a & L_b & 1\\
J_b & J_f & S \\
\end{array}
\right\rbrace
\left\lbrace
\begin{array}{ccc}
L_a & L_b & 1\\
J_{b^\prime} & J_{f^\prime} & S \\
\end{array}
\right\rbrace
\nonumber \\ && \!\!\!\!\!\!\!\!\!\! \times
\left\lbrace
\begin{array}{ccc}
L_a & L_b & 1\\
J_{b^{\prime\prime}} & J_a & S \\
\end{array}
\right\rbrace
\left\lbrace
\begin{array}{ccc}
L_a & L_b & 1\\
J_{b^{\prime\prime\prime}} & J_{a^\prime} & S \\
\end{array}
\right\rbrace
\nonumber \\ && \!\!\!\!\!\!\!\!\!\! \times
(-1)^Q \mathcal{T}^K_{-Q}(i,{\bm n})
\mathcal{T}^{K^\prime}_Q(j,{\bm n}^\prime)\ .
\label{final-rm}
\end{eqnarray}
The assumptions underlying the derivation of Equation~(\ref{final-rm}) are that 
the lower levels are unpolarized and infinitely sharp. 
See \citet{sow14} for details on the terminology and the derivation.

\section{Single scattering polarization with PBE}
\label{results}
As an example to study the PBE in fine structure states
we consider the $L=0$ and $L=1$ terms of the two stable isotopes of
neutral lithium, namely $^7$Li and $^6$Li. The isotopic shifts are measured
with respect to the reference isotope $^7$Li. In our calculations we use the
isotopic shift values given in Table 1 of \citet{bel09}. The abundances for the
two isotopes are also read from the same table.
The total electron spin, $S=1/2$. The coupling between $\bm L$ and $\bm S$ results
in $J=3/2$ and $1/2$ for the $L=1$ term and $J=1/2$ for the $L=0$ term.
The transitions between these $J$ states in the absence of magnetic fields
results in the D$_1$ and D$_2$ lines (obeying the selection rules
$\Delta S=0,\ \Delta J=0,\pm1$). The wavelengths of these transitions are listed in
Table \ref{tab-0}.
In the presence of a magnetic field, the non-degenerate magnetic substates give
rise to 10 allowed transitions (according to the selection rule $\Delta\mu=0,\pm1$)
in each of the two isotopes. Among these 10 transitions,
6 are between the magnetic substates of the upper $J=3/2$ and the lower $J=1/2$
states and the rest are between those of the upper $J=1/2$ and the lower 
$J=1/2$ states. These transitions can be classified into three
groups: $\sigma_{\rm r}$ ($\Delta\mu=-1$),
$\pi$ ($\Delta\mu=0$), and $\sigma_{\rm b}$
($\Delta\mu=+1$). Note that $\Delta\mu=\mu_b-\mu_a$
where $\mu_b$ are the magnetic substates of the upper $J$ state and $\mu_a$ are the
magnetic substates of the lower $J$ state. The magnetic ($\pi$ and $\sigma$)
components of the {\rm D}$_1$ lines
will be denoted with a prime in the following discussions
for the sake of clarity and distinction. As per this classification, the
{\rm D}$_2$ line gives rise to two $\sigma_{\rm r}$, two $\pi$, and two $\sigma_{\rm b}$
components while the {\rm D}$_1$ line gives rise to one $\sigma^\prime_{\rm r}$,
two $\pi^\prime$, and one $\sigma^\prime_{\rm b}$ components, in each of the two
isotopes. These are tabulated in Table~\ref{tab-1}. The magnetic components of the
two isotopes will be distinguished by their mass numbers indicated in the superscripts
to the $\pi$ and $\sigma$ components. For the computation of
the Stokes profiles presented in 
Figures~\ref{sz-1}--\ref{cont}, we assume that an
unpolarized radiation is incident on the atom at an angle ${\cos}\ \theta^\prime=1$
and gets scattered in a direction ${\rm cos}\ \theta=0$, where $\theta^\prime$ and
$\theta$ are the colatitudes. The values of the azimuths $\chi$ and $\chi^\prime$
for scattered and incident rays, respectively, are assumed
to be zero in this single 90\textdegree\ scattering event.
The scattered ray so obtained is given by the first column of the RM, which 
is then integrated over the incoming frequencies to get the singly scattered 
Stokes profiles. For the Li\,{\sc i} D line system, the Stokes parameters are obtained
by linearly combining the Stokes parameters computed for the individual isotopes
weighted by their respective abundances. Such a linear superposition is allowed
because the lines are optically thin.
Following \citet{bel09} we use a Doppler width of 60 m\AA\ for all the components.
\begin{table}[ht]
 \begin{centering}
  \begin{tabular}{|c|c|c|c|}
   \hline
Isotope & Line & $\lambda$ (\AA) & $A$ (s$^{-1}$)\\
\hline
$^6$Li & D$_1$ & 6708.05534 & 3.689 $\times10^7$\\
$^6$Li & D$_2$ & 6707.90232 & 3.689 $\times10^7$\\
$^7$Li & D$_1$ & 6707.89719 & 3.689 $\times10^7$\\
$^7$Li & D$_2$ & 6707.74416 & 3.689 $\times10^7$\\
\hline
  \end{tabular}
\caption{Wavelengths and Einstein $A$ coefficients for the D line transitions
of Li isotopes.}
\label{tab-0}
\end{centering}
\end{table}

\begin{table*}[ht]
 \begin{centering}
\begin{tabular}{|c|c|cc|cccc|}
\hline
$J_a\diagdown J_b$ &  & 1/2 &  & 3/2 & & & \\
\hline
\ \ \ & $\mu_a\diagdown \mu_b$ & -1/2 & +1/2 & -3/2 & -1/2 & +1/2 & +3/2\\
\hline
1/2 &-1/2 & $\pi^\prime$ & $\sigma^\prime_{\rm b}$ & $\sigma_{\rm r}$ & $\pi$ &
$\sigma_{\rm b}$ & NA\\
 & +1/2 & $\sigma^\prime_{\rm r}$ & $\pi^\prime$ & NA & $\sigma_{\rm r}$ &
$\pi$ & $\sigma_{\rm b}$\\
\hline
\end{tabular}
\caption{The list of transitions between the magnetic substates of the upper and 
the lower $J$ states. NA - Not Allowed. The magnetic components of the two isotopes
are distinguished in the following by their atomic masses indicated in the superscripts.}
\label{tab-1}
\end{centering}
\end{table*}

\subsection{The diagonalization procedure}
The non-zero matrix elements of the total Hamiltonian defined in 
Equation~(\ref{mat-diag}) are of the form given by
Equations (3.61a) and (3.61b) of LL04. Following \citet{landi78} we write a program
to diagonalize the total Hamiltonian. The numerical diagonalization is performed
using the Givens-Householder method. We obtain the
eigenvalues in terms of the energy shifts
from the parent $L$ state and the eigenvectors in terms of the $C$-coefficients.
By making use of these energy shifts, we determine the energies of the
$L_a=0$ and $L_b=1$ terms. Since $J$ is not a good
quantum number in the PB regime, we cannot use either {\rm D}$_1$ or {\rm D}$_2$
wavelengths. For the atomic system we have considered, the line center
wavelengths correspond to the transitions
$^7L_a=0$ \textrightarrow $^7L_b=1$ and $^6L_a=0$ \textrightarrow $^6L_b=1$,
which are, respectively, 6707.79517\,\AA\ and 6707.95333\,\AA.

In the presence of a magnetic field, the degeneracy of the magnetic substates is lifted
and the spectral lines split into magnetic components. It is possible to obtain the
magnetic shifts and strengths of these components by making use of
the $C$-coefficients and the energy eigenvalues. 
The normalized strengths of the transitions which connect the magnetic substates of 
the lower term ($L_aS$) with those of the upper term ($L_bS$) are given by 
\begin{eqnarray}
 && \mathcal{S}_q^{j_a\mu_a,j_b\mu_b}={\alpha}
\sum_{J_aJ_{a^\prime}J_bJ_{b^\prime}}\frac{3}{2S+1}
\nonumber \\ &&
\times C^{j_a}_{J_a}(L_aS,\mu_a) C^{j_a}_{J_{a^\prime}}(L_aS,\mu_a)
\nonumber \\ &&
\times C^{j_b}_{J_b} (L_b S,\mu_b)
C^{j_b}_{J_{b^\prime}}(L_b S,\mu_b) (2L_a+1)
\nonumber \\ &&
\times\sqrt{(2J_a+1)(2J_{a^\prime}+1)(2J_b+1)(2J_{b^\prime}+1)}
\nonumber \\ &&
\times
\left (
\begin{array}{ccc}
J_b & J_a & 1\\
-\mu_b & \mu_a & -q \\
\end{array}
\right )
\left (
\begin{array}{ccc}
J_{b^\prime} & J_{a^\prime} & 1\\
-\mu_b & \mu_a & -q \\
\end{array}
\right )
\nonumber \\ && \times
\left\lbrace
\begin{array}{ccc}
L_a & L_b & 1\\
J_b & J_a & S \\
\end{array}
\right\rbrace
\left\lbrace
\begin{array}{ccc}
L_a & L_b & 1\\
J_{b^\prime} & J_{a^\prime} & S \\
\end{array}
\right\rbrace\ .
\label{str}
\end{eqnarray}
Here $\alpha$ represents the percentage abundance of the isotope.
The magnetic shifts are given by
\begin{eqnarray}
\!\!\!\!\!\!\!\!\!\!\!\!\!\!\!
&& \Delta^{j_aj_b}_{\mu_a\mu_b}=\frac{E_{j_b}
(L_b S,\mu_b)-E_{j_a}(L_aS,\mu_a)}{h} + {\delta E_{\rm iso}}\ ,
\nonumber \\ \!\!\!\!\!\!\!\!\!\!\!\!\!\!\!&& 
\label{zesh}
\end{eqnarray}
where $E_j$ are the energy eigenvalues. $h$ is the Planck's constant.
$\delta E_{\rm iso}$
is the isotopic shift measured with respect to the reference isotope $^7$Li.
Note that $\delta E_{\rm iso}$ is zero for the reference isotope $^7$Li.
$\Delta$s are given in frequency units.
\begin{figure}[ht]
\begin{center}
 \includegraphics[scale=0.45]{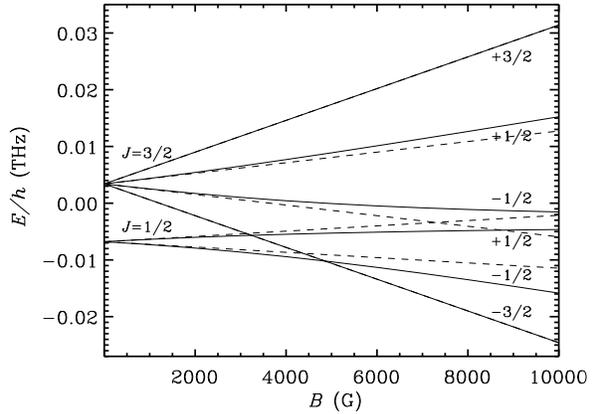}
\end{center}
\caption{Level crossing between the magnetic substates belonging to the $^2{\rm P}$ term
of $^7$Li isotope in the presence of a magnetic field. A comparison between the
splittings produced by including (solid lines) and neglecting (dashed lines) PBE.
The $^2{\rm P}$ term of $^6$Li isotope gives a similar diagram except for the
isotopic shift.}
\label{level-fig}
\end{figure}

Figure~\ref{level-fig} shows the behavior of the energy eigenvalues of the magnetic
substates belonging to upper $J$ states as a function of the magnetic field
strength $B$. As described in LL04, we define a parameter
$\gamma$ as
\begin{eqnarray}
 && \gamma=\frac{\mu_0B}{\zeta}\ ,
\label{gam}
\end{eqnarray}
which is a ratio of the magnetic energy to the fine structure energy.
The energy eigenvalues diverge linearly with increase in the magnetic field
strength for $\gamma\ll1$.
This regime is called the linear Zeeman regime. For intermediate values of $\gamma$,
non-linearity sets in and the eigenvalues start to cross. This regime is called the
incomplete PB regime. For $\gamma\gg1$ the eigenvalues again vary linearly with
$B$ and this regime is called the complete PB regime. For the atomic system considered
we see two level-crossings. The values of $\gamma$ and $B$ for which these crossings
occur are listed in Table \ref{tab-2}.

The solid lines in Figure~\ref{level-fig} are computed taking
the non-zero non-diagonal elements
of the total Hamiltonian (see Equation~(3.61b) of LL04) 
into account while the dashed lines are computed by neglecting them.
This means that, for the dashed lines, the splitting produced by the magnetic field
is just given by the expression $\mu_0Bg_{J_b}\mu_b$
(where $g_{J_b}$ is the Land\'e factor).
We can clearly see the differences that
PBE makes to the energy eigenvalues, from this diagram. The substates with $\mu_b=+3/2$
and $-3/2$ show the same behavior irrespective of whether PBE is included or not. This is
because the contribution from the non-diagonal elements for these $\mu_b$s are zero,
as these $\mu_b$s can be assigned to only $J_b=3/2$ state. For
the other magnetic substates the splitting becomes nonlinear because of the contribution
from the non-diagonal elements to the total splitting caused by the magnetic field.
In particular we notice that the magnetic substates which cross in the case of Zeeman
effect avoid crossing one another in the case of PBE. For example, the 
$\mu_{b^\prime}=+1/2$ belonging to $J_{b^\prime}=1/2$ and $\mu_b=-1/2$
belonging to $J_b=3/2$ cross at $B \sim 7.3$\,kG 
when magnetic splittings are computed using the Zeeman effect. On the other hand, 
when PBE is included to compute the magnetic splitting, these substates do not 
cross. This is known as avoided crossing
(also known as anti-level-crossing). As a consequence of this we find that the
polarization in the asymptotic limit of $B\rightarrow\infty$ is larger than that when
$B\rightarrow0$. See \citet{bom80} and LL04 for more details on this effect.
\begin{table}[ht]
 \begin{centering}
\begin{tabular}{|c|c|cc|}
\hline
$\mu_{b^\prime}$ & $\mu_b$ & $\gamma$ & $B$ (kG)\\
\hline
-3/2 & 1/2 & 0.667 & 3.238\\
-3/2 & -1/2 & 1.0 & 4.855\\
\hline
\end{tabular}
\caption{The values of $\gamma$ and approximate values of $B$ at level-crossings
for the $^2{\rm P}$ term
of $^7$Li for which the level-crossing diagram shown in Figure~\ref{level-fig} is made.
The same table holds good for the $^2{\rm P}$ term of $^6$Li.}
\label{tab-2}
\end{centering}
\end{table}
\begin{figure*}
\begin{center}
 \includegraphics[scale=0.39]{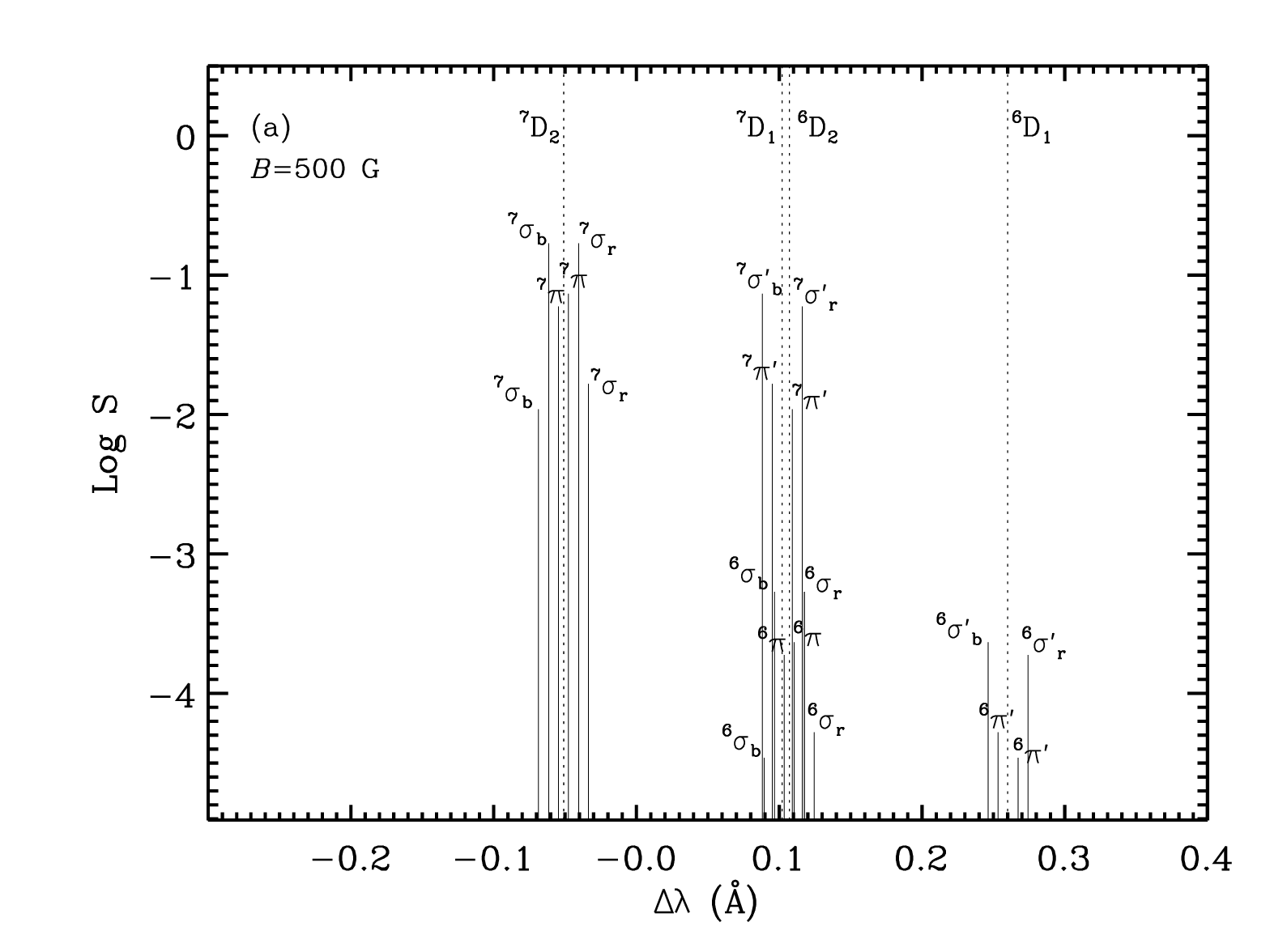}
 \includegraphics[scale=0.39]{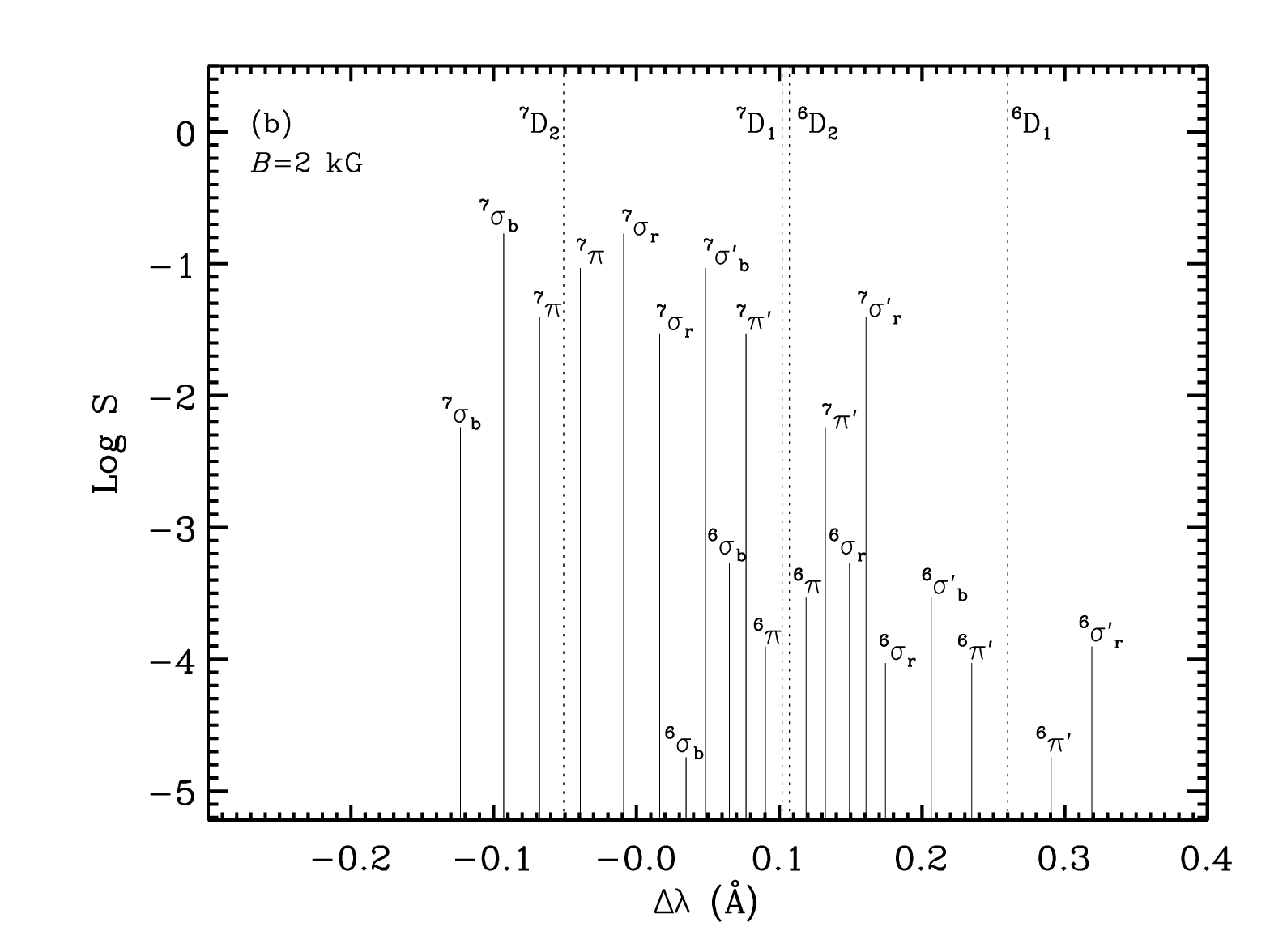}
 \includegraphics[scale=0.48]{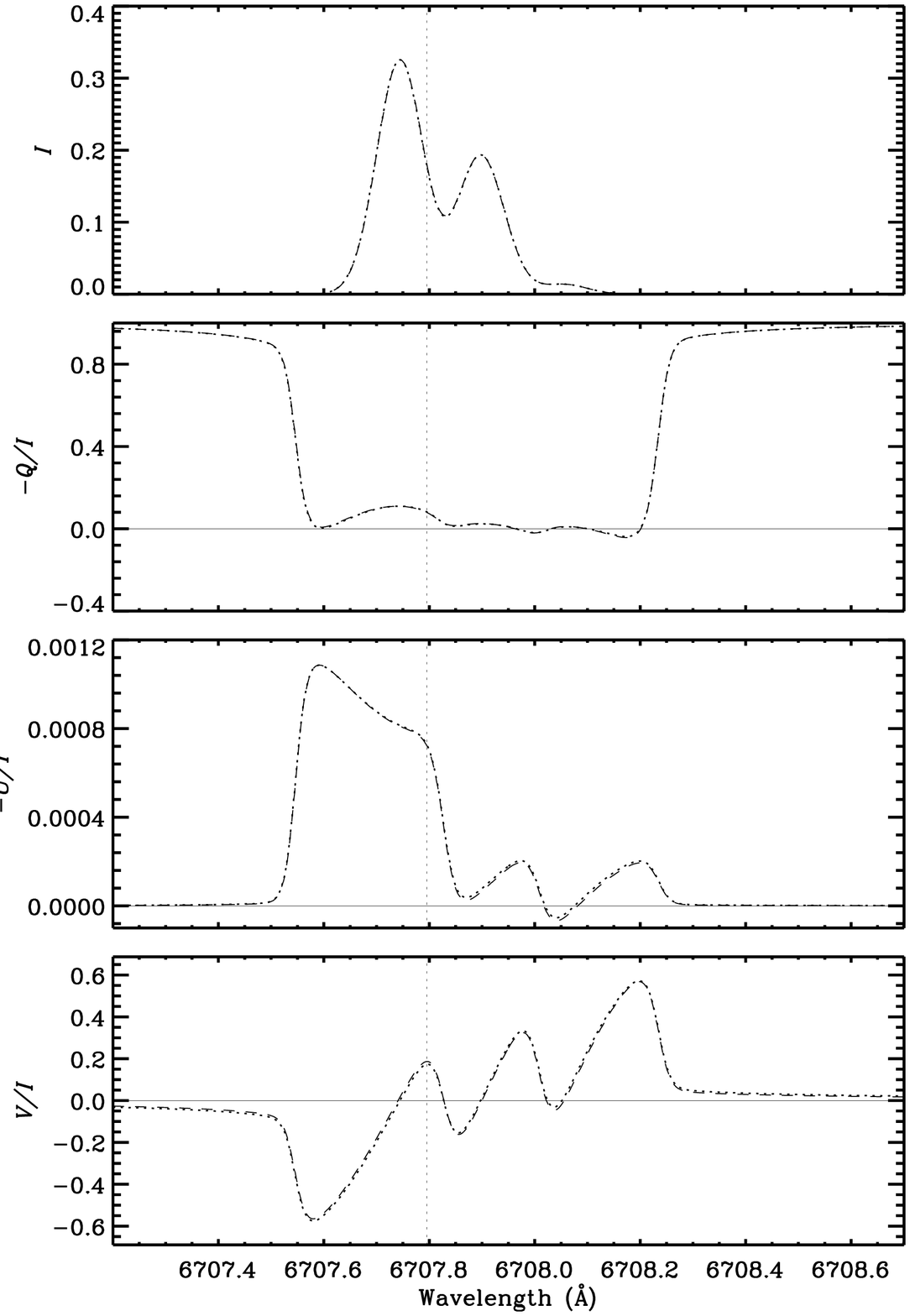}
 \includegraphics[scale=0.48]{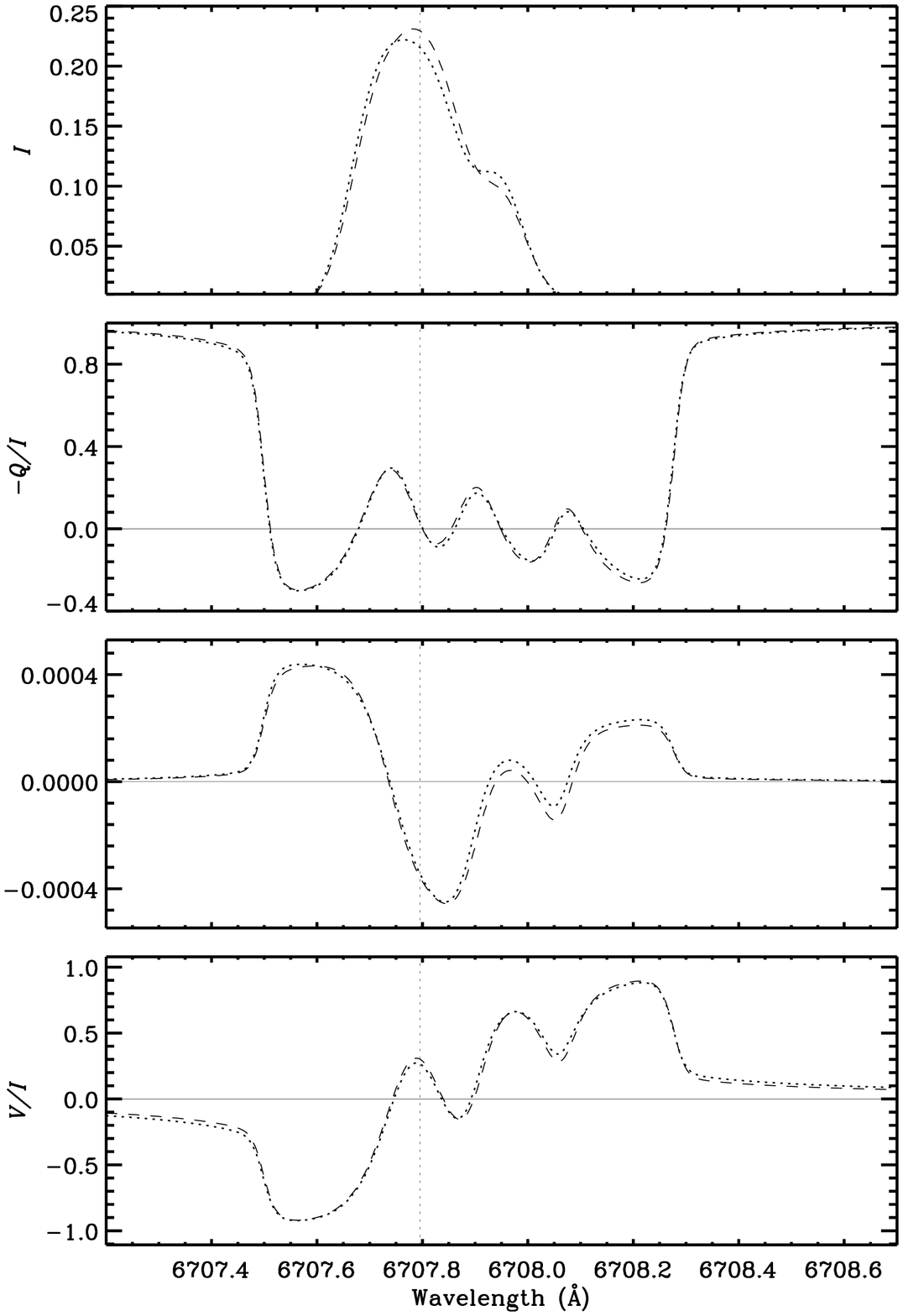}
\end{center}
\caption{Top panels refer to the line splitting diagrams (line strength $\mathcal{S}$
vs. wavelength shift). The other panels show the 
comparison of the Stokes profiles computed using the PB-FS code (dashed lines)
with the Zeeman-FS code (dotted lines). The two columns correspond to different field
strengths as indicated in the line splitting diagrams. The vertical dotted lines
in the line splitting diagrams indicate the positions of the $^7$Li and $^6$Li D lines.
The vertical dotted lines in other panels correspond to the line center
wavelength of the $L=0\rightarrow1\rightarrow0$ transition in the reference isotope
$^7$Li. The orientation of the magnetic field is given by
$(\theta_B, \phi_B)=(90\degree, 45\degree)$.}
\label{sz-1}
\end{figure*}
\begin{figure*}
\begin{center}
 \includegraphics[scale=0.39]{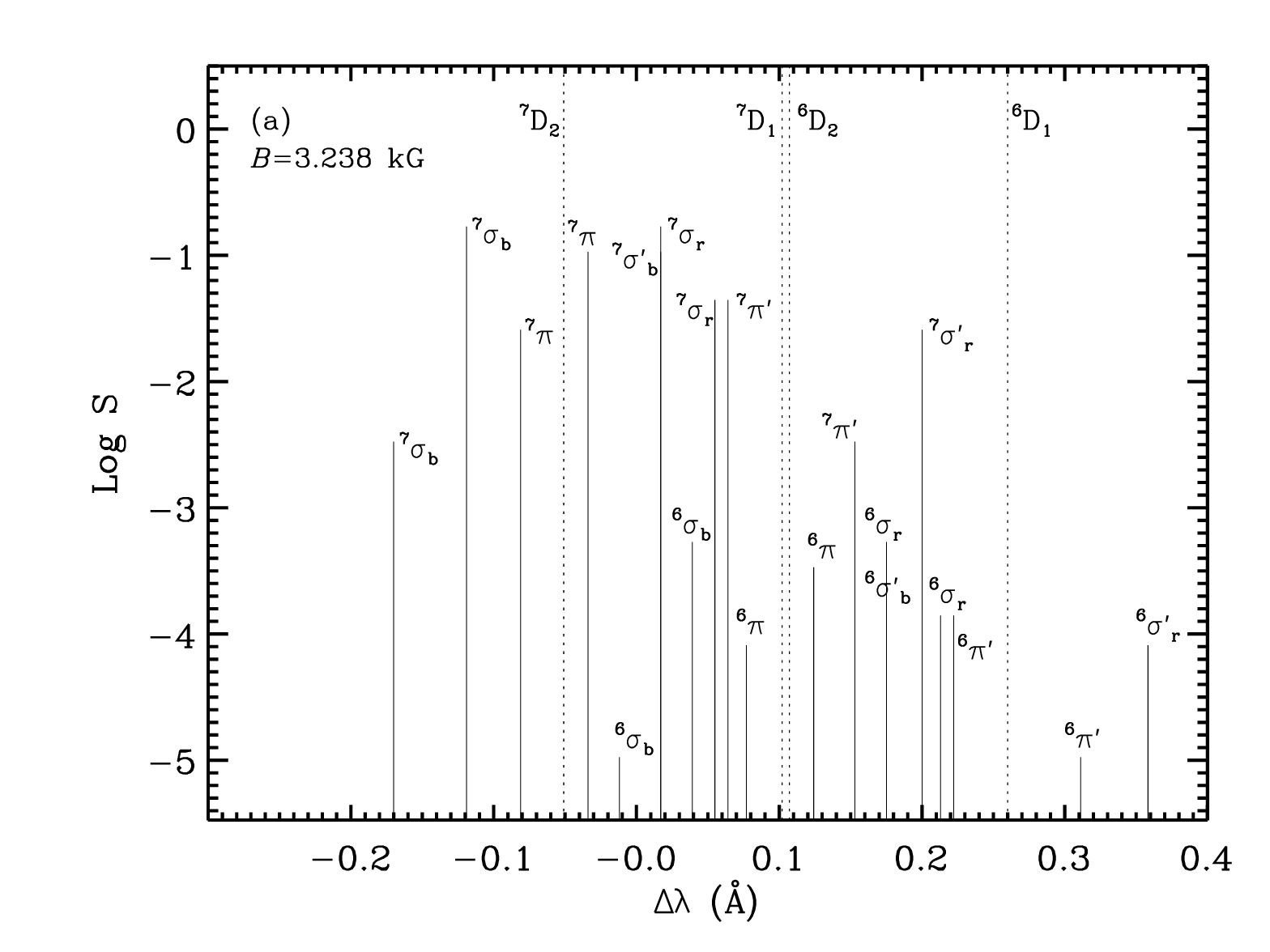}
 \includegraphics[scale=0.39]{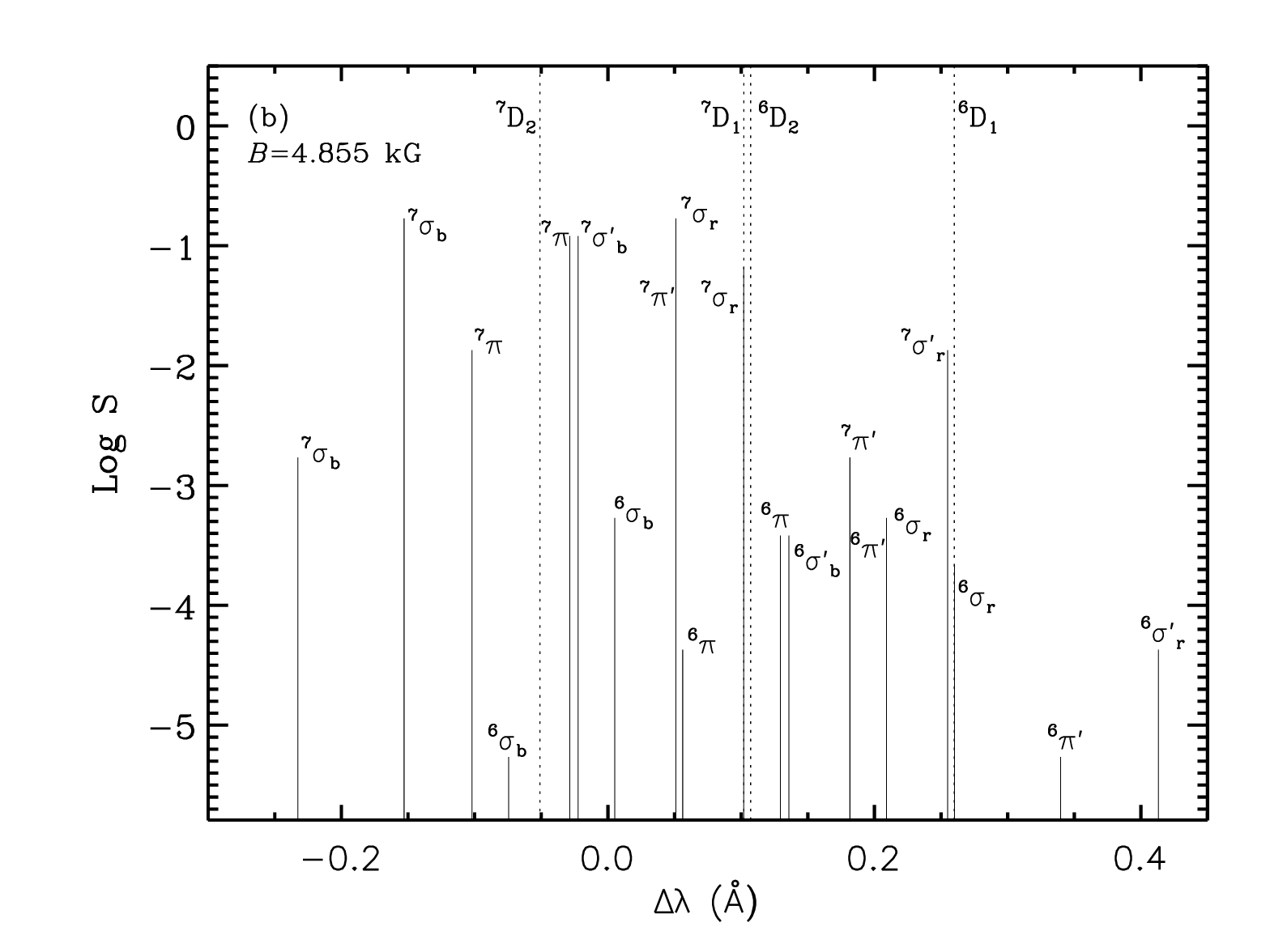}
 \includegraphics[scale=0.48]{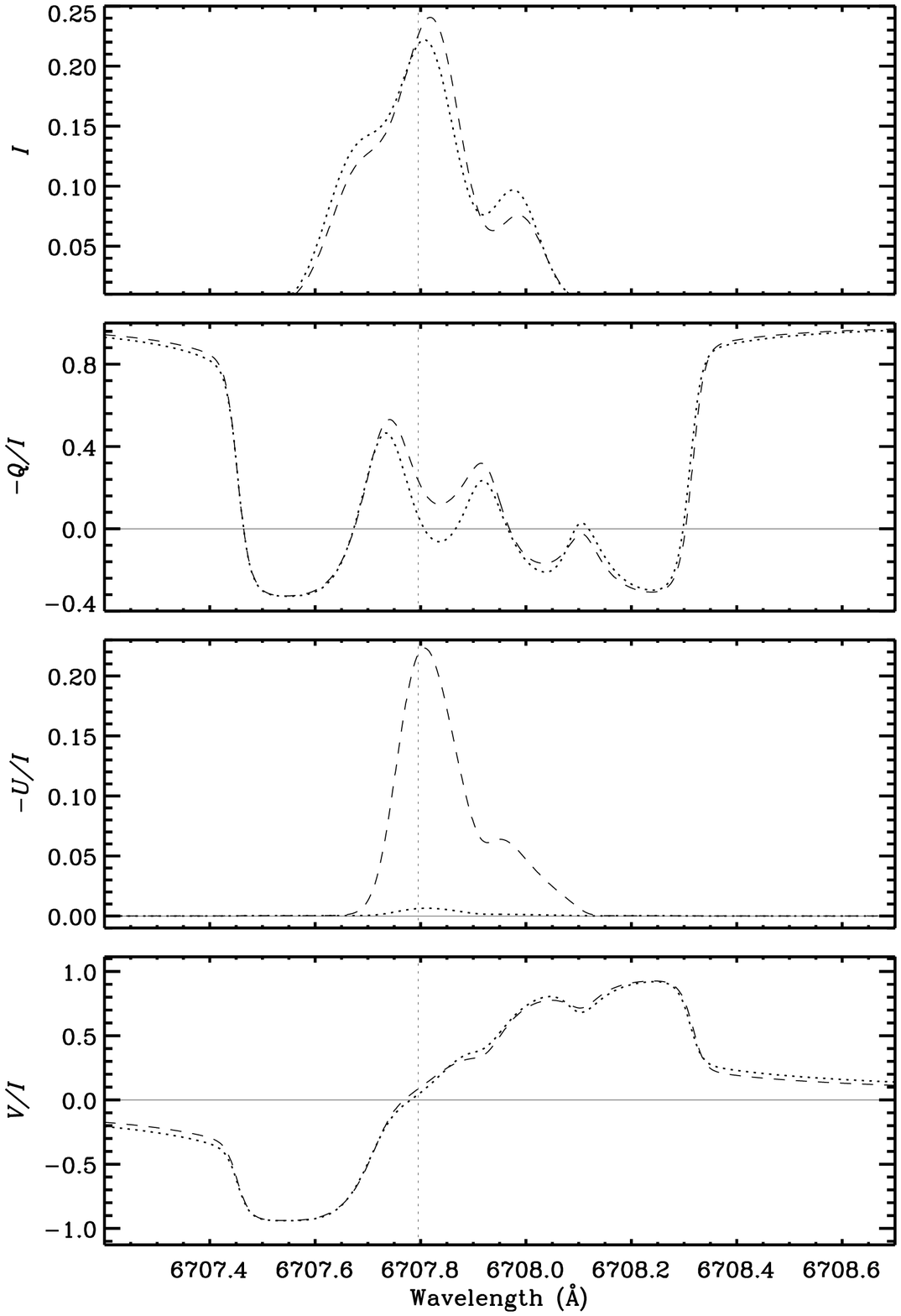}
 \includegraphics[scale=0.48]{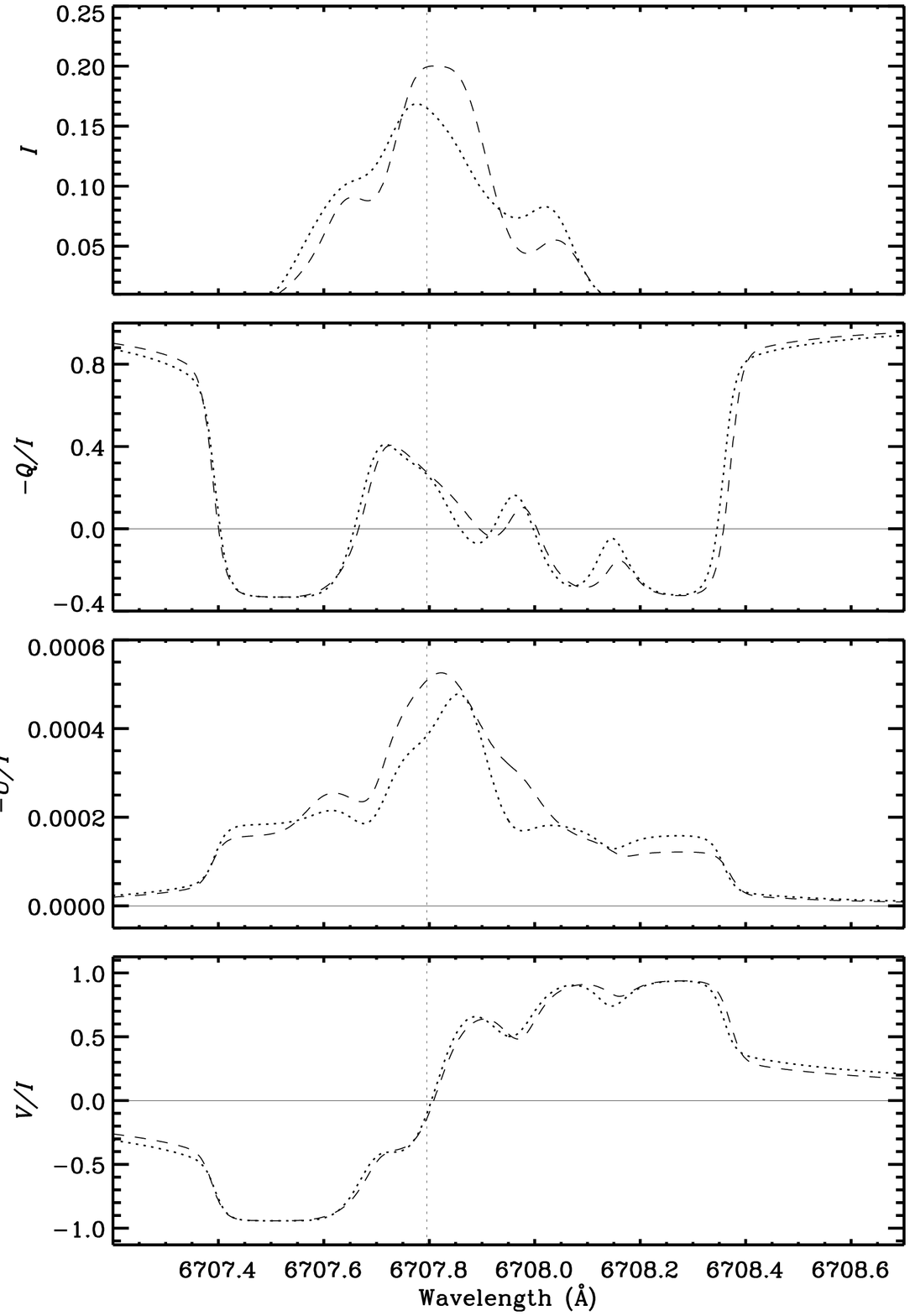}
\end{center}
\caption{Same as Figure~\ref{sz-1} except for field strengths.}
\label{sz-2}
\end{figure*}

\subsection{Comparison of Stokes profiles computed using linear Zeeman and PB effects}
In the linear Zeeman regime, the RM presented in Section~\ref{pbl}
reduces to Equation~(25) of \citet{smi11a}. In order to show the effects of level-crossing
we compare the results of our code which programs Equation~(\ref{final-rm}) 
(hereafter called as PB-FS code) with the results of \citet{smi11a} (hereafter called as
Zeeman-FS code). This comparison is shown in Figures~\ref{sz-1} and \ref{sz-2}.
The Stokes profiles from the two codes match very well up to 500 G for which
$\gamma=0.1029$. According to the classification scheme discussed in the previous section,
we are still in the linear Zeeman regime for this field strength. For field strengths
larger than 500 G, the differences start to appear as we already enter the
non-linear regime in which the linear Zeeman approximation (Zeeman-FS code) breaks down.
The separation between the magnetic components (which increases with an increase in $B$)
is no longer given by $\mu_0Bg_{J_b}\mu_b$. Hence there is a difference in the line center
positions of the magnetic components computed from the two codes. These small differences
are clearly seen in intensity ($I$) profiles (see right panels of Figure~\ref{sz-1}).
For level crossing field strengths (3.238 kG and 4.855 kG) the Stokes profiles computed
from the Zeeman-FS and PB-FS codes differ drastically.
The Zeeman-FS code therefore does not cover all the field strength ranges that we
can expect on the Sun.


\subsection{Stokes profiles in the PB regime}
By making use of the strengths and shifts of the PB components obtained from
the diagonalization code, we have made line splitting diagrams where the log of the
PB component strengths ($\mathcal{S}$) given in Equation~(\ref{str})
are plotted against their wavelength shifts from the line center computed
using Equation~(\ref{zesh}).
These are shown in Figures~\ref{sz-1} and \ref{sz-2}
along with the Stokes profiles for different values of $B$. 
The zero on the x-axis of the line splitting diagram corresponds to the line center
wavelength of the $L=0\rightarrow1$ transition in $^7$Li.

For $B=500$ G (linear regime), the magnetic components are
separated into two bunches of 6 and 4 each, in both the isotopes. The magnetic components
of the $^7$D$_1$ and $^6$D$_2$ lines superpose due to their proximity in wavelength.
The splitting is more or less symmetric about the line centers of the
{\rm D}$_1$ and {\rm D}$_2$ lines but the strengths of the components vary depending 
on the values of the magnetic quantum numbers $\mu_a$ and $\mu_b$ (see the left top panel 
of Figure~\ref{sz-1}). The same is reflected
in the intensity profiles. The three peaks seen
in intensity correspond to the three bunches of magnetic components with the amplitudes
of the peaks being proportional to the relative abundances of the two isotopes.
The $Q/I$ and $U/I$ profiles show typical signatures of Hanle effect especially at the
position of the $^7$D$_2$ line, namely a depolarization of the $Q/I$ with respect to
the non-magnetic value (0.428 in the line core) and a generation of $U/I$ signal.
The $^7$D$_1$ and $^6$D$_1$ lines are non-polarizing
and hence are unaffected by Hanle effect. The
$^6$D$_2$ line, although affected by Hanle effect, produces insignificant signatures due
to its small abundance. In spite of these we see $U/I$ peaks at the positions
corresponding to ($^7$D$_1$,$^6$D$_2$) and $^6$D$_1$, the origin of which is not clear.
They are possibly due to the interference between the D lines.
Note however that these $U/I$ signatures are about three orders of magnitude smaller
than the corresponding $Q/I$ signatures and are therefore much too weak to be observable.
The $V/I$ arises due to the longitudinal component of the magnetic field.

For $B=2$ kG (see panels (b) of Figure~\ref{sz-1}), the components are well 
separated and their strengths change because
of the dependence on the $C$-coefficients which vary with $B$. The components cannot be
resolved in intensity as their Doppler width is larger than the separation between them. 
The decrease in the intensity is due to an increased separation between the magnetic
components with increasing magnetic field strength. In $Q/I$, a three-lobed Zeeman like
pattern is seen, $U/I$ is very small because of the geometry. The
$V/I$ profiles become broader as expected. The $\sigma_{\rm b}$ components show opposite
behavior to those of $\sigma_{\rm r}$ again as expected. Positive peaks appear at
the positions of the $\sigma_{\rm r}$ components while negative peaks occur 
at the positions corresponding to $\sigma_{\rm b}$.

\begin{figure*}
 \begin{center}
  \includegraphics[scale=0.7]{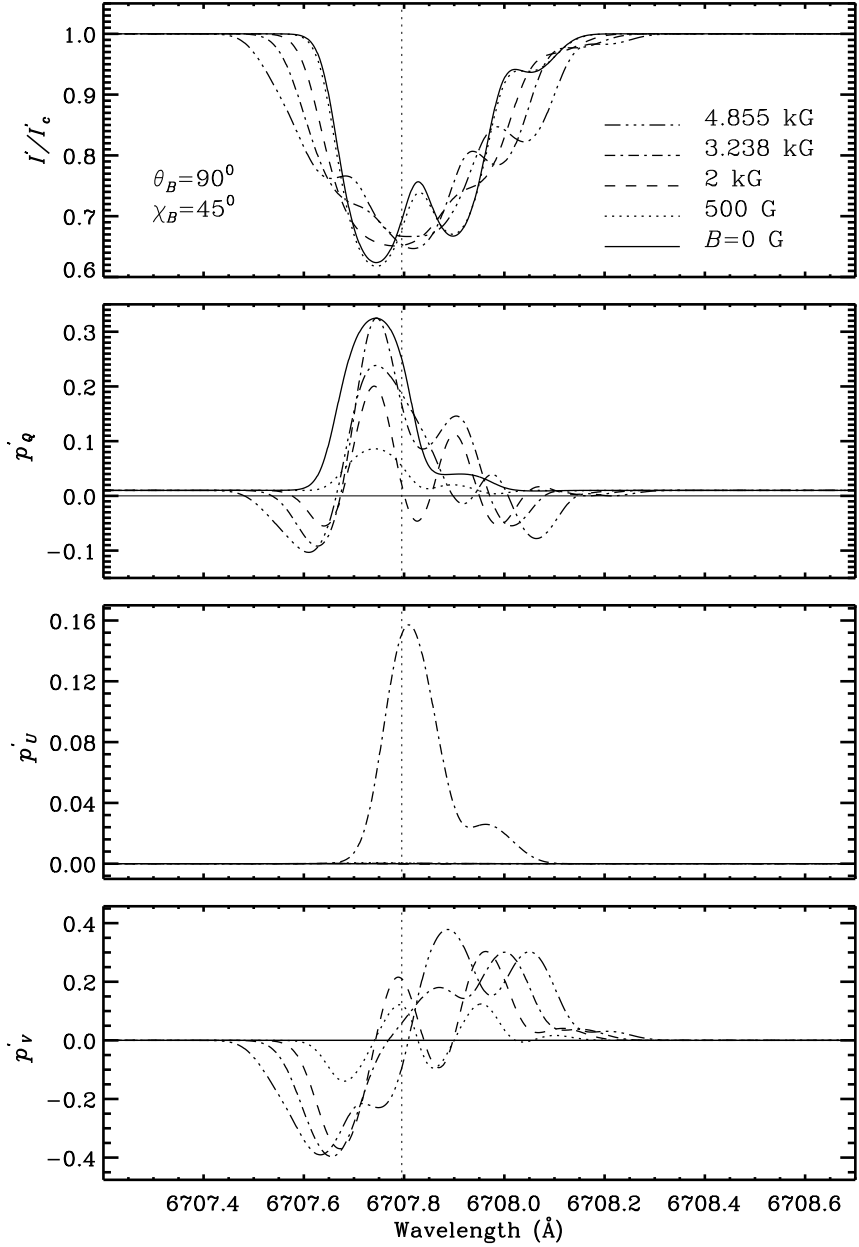}
 \end{center}
\caption{Stokes profiles computed using PB-FS code by including the contribution
from the polarized continuum. The line types and the orientation of the magnetic field
are indicated in the intensity panel. The vertical dotted lines in all the panels
correspond to the line center wavelength of the $L=0\rightarrow1\rightarrow0$ transition
in the reference isotope $^7$Li. See text for more details.}
\label{cont}
\end{figure*}

For $B=3.238$ kG (see panels (a) of Figure~\ref{sz-2}) at which the first level crossing 
occurs, we see the interference between the $\sigma_{\rm r}$ and $\sigma^\prime_{\rm b}$
components in the two isotopes. Their positions overlap as can be seen from the line
splitting diagram. Interestingly, at these positions we see a generation of $U/I$ signal
due to interference between the magnetic substates (Hanle effect). The $V/I$ signals of
the D lines overlap giving rise to a broader profile.

For $B=4.855$ kG (see panels (b) of Figure~\ref{sz-2}) at which the second level 
crossing occurs, there is interference between the $\sigma_{\rm r}$ and $\pi^\prime$
components in the two isotopes. The $U/I$ signal is generated due to the Hanle effect.

In Figure~\ref{cont} we show the Stokes profiles obtained from the PB-FS code in
the presence of a weakly polarized background continuum. The contribution from the
continuum is included in the same way as in \citet{s98}. For the sake of clarity, we
recall his Equations (58) and (61):
\begin{eqnarray}
 && I^\prime/I^\prime_c=1-\beta+\frac{a}{I+a}\beta
\label{cont-int}
\end{eqnarray}
\begin{eqnarray}
 && p^\prime=\frac{I}{I+a}p+\frac{a}{I+a}b
\label{cont-pol}
\end{eqnarray}
In the above equations $I$ and $p$ are the intensity and fractional polarization
given by $-Q/I,-U/I$ and $V/I$ in the absence of the continuum. The corresponding 
quantities in the presence of the continuum are $I^\prime$ and $p^\prime$. 
The limb-darkening parameter, $\beta$, and the continuum strength parameter, $a$,
are chosen as 0.5 and 0.1, respectively. Such a large value of $a$ is chosen
to make the Stokes profiles resemble closely with the profiles presented for the
non-magnetic case in \citet{bel09}. The continuum polarization parameter, $b$,
is chosen as 0.01 for $Q$ and 0 for $U$ and $V$. With this choice we obtain
profile shapes of the kind that we expect in the Sun's spectrum.
In particular our non-magnetic $p^\prime_Q$ profile (solid line) resembles the shape
of the corresponding profile observed with ZIMPOL \citep{s11}. The intensity
profiles appear as broad absorption lines. The fractional linear polarization 
$p^\prime_Q$ ($=-Q^\prime/I^\prime$) approaches the continuum polarization value ($b=0.01$)
away from the line center. The $p^\prime_U$ ($=-U^\prime/I^\prime$) and 
$p^\prime_{V}$ 
($=V^\prime/I^\prime$) profiles retain their overall shape compared to the 
pure line case without continuum, although
their amplitudes decrease because of the contribution from the continuum 
strength parameter $a$ to $I^\prime$.
As can be seen from the figure, the shape of the $p^\prime_{Q}$ profile for the zero field
case (solid line) compares well with the corresponding profiles presented in
\citet{bel09}. Note that since the Stokes profiles are computed here for a single
scattering event, only the shape and not the amplitude is comparable to the corresponding
profiles presented by \citet{bel09}.

\subsection{Polarization diagrams}
\begin{figure}[ht]
 \begin{center}
  \includegraphics[scale=0.9]{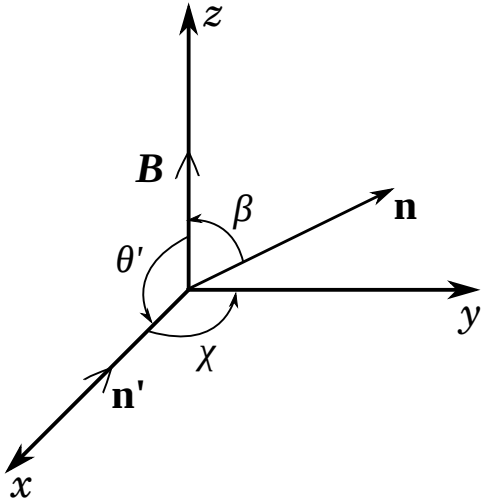}
 \end{center}
\caption{Geometry considered for computing the polarization diagrams shown in Figure
\ref{2p-pol}. $\beta$ is the angle between
the magnetic field vector and the scattered beam. The incident radiation is characterized
by $(\theta^\prime,\chi^\prime)=(90\degree,0\degree)$ and the scattered radiation by
$(\theta,\chi)=(\beta,90\degree)$. The magnetic field inclination, $\theta_B=0\degree$
and its azimuth, $\chi_B=0\degree$ (magnetic reference frame).}
\label{geom}
\end{figure}

\begin{figure*}
\begin{center}
 \includegraphics[scale=0.6]{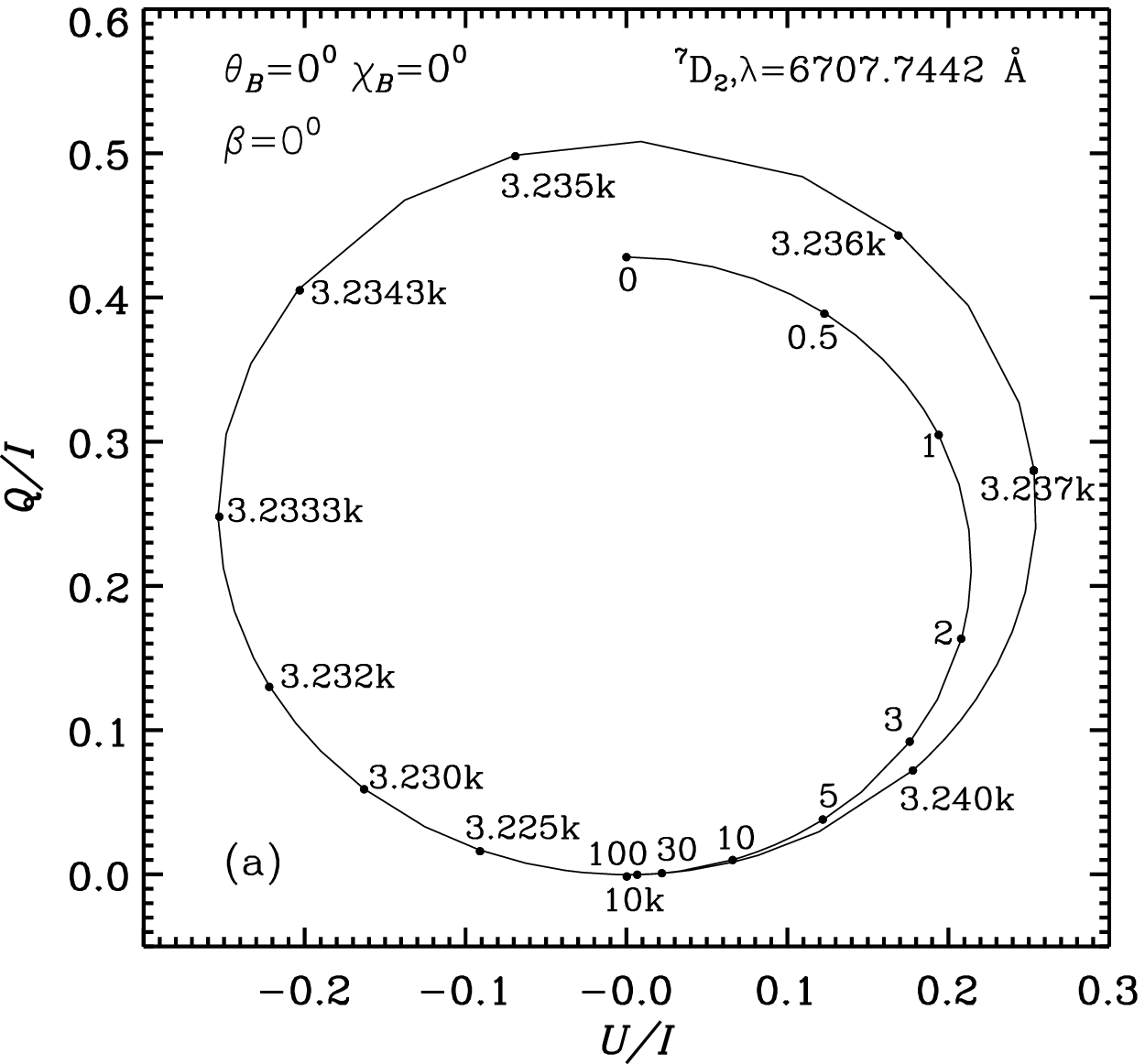}
 \includegraphics[scale=0.6]{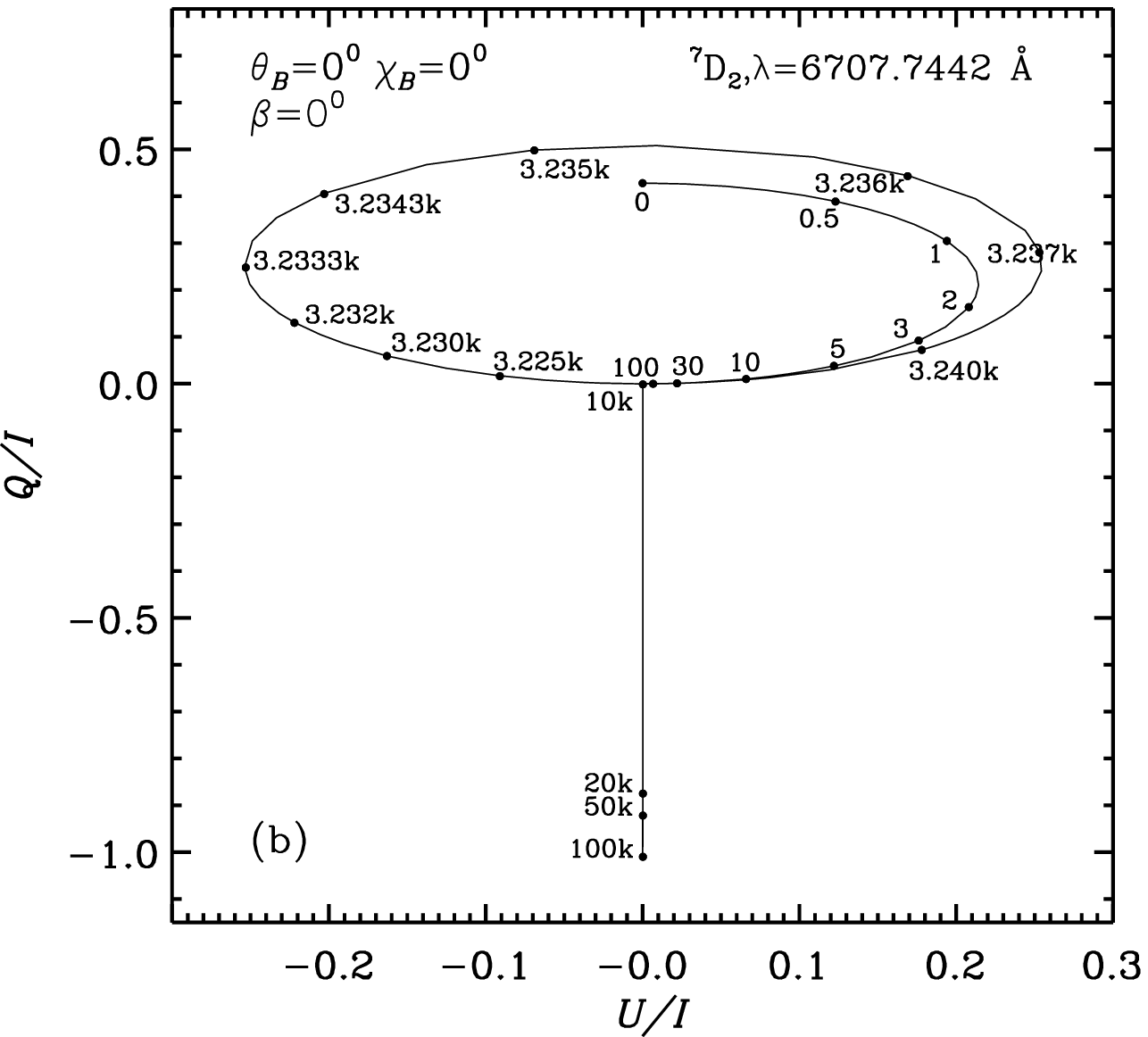}
 \includegraphics[scale=0.6]{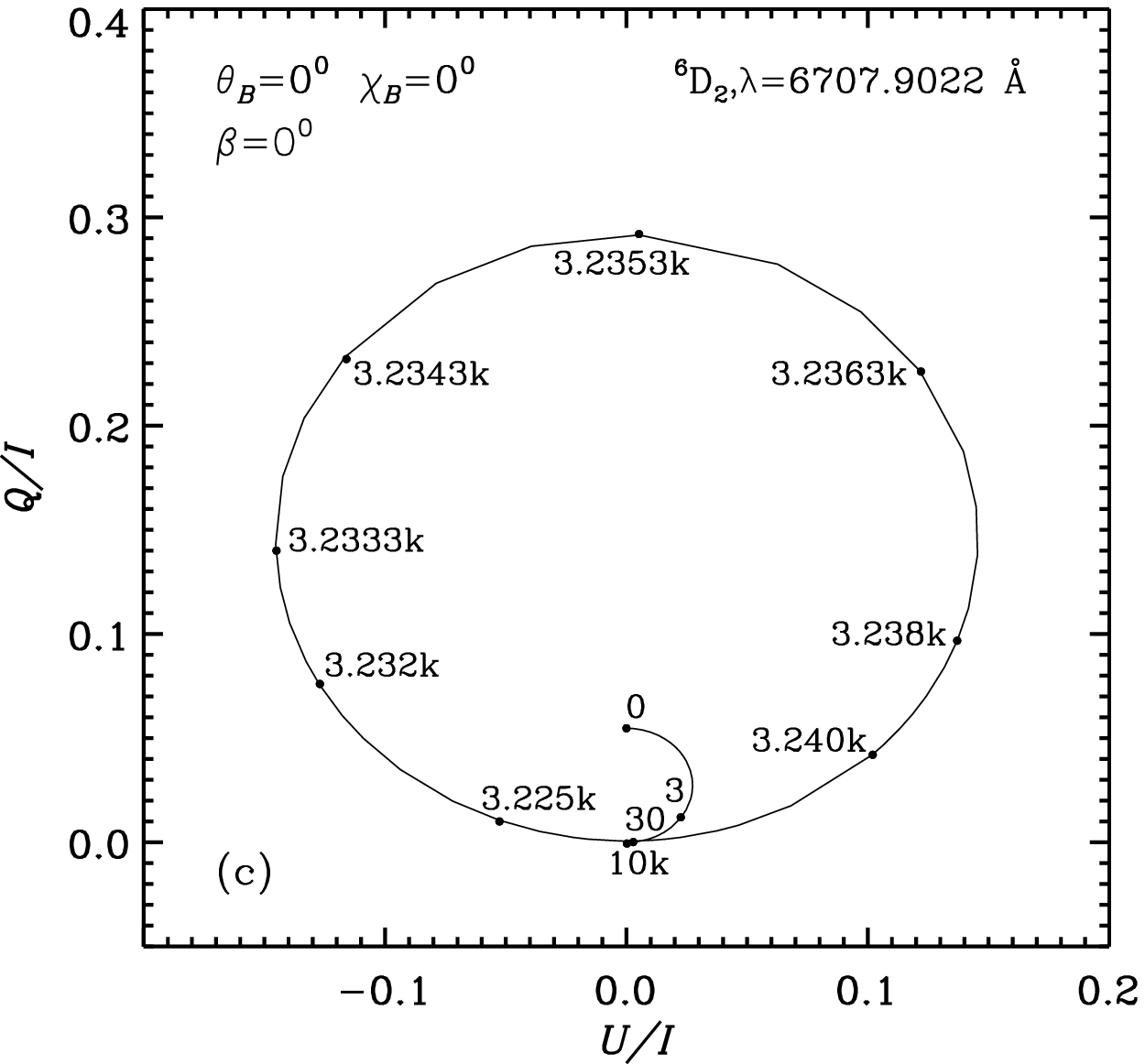}
 \includegraphics[scale=0.6]{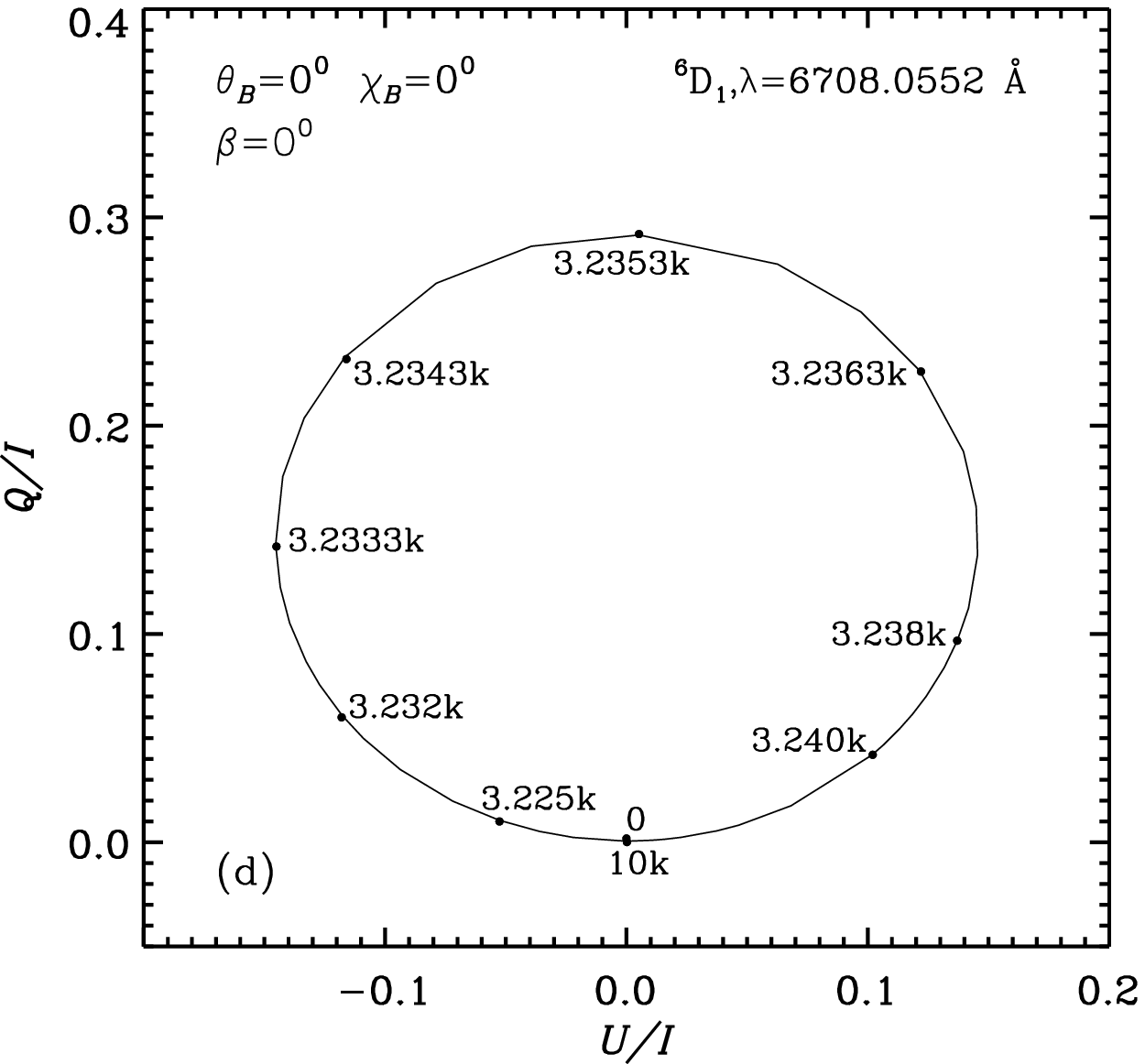}
\end{center}
\caption{Polarization diagrams computed using the PB-FS code at the line
center wavelengths of the $^7$Li and $^6$Li D lines. The
curves are marked by the values of the magnetic field strength $B$ in G.
A `k' means a factor of 1000.}
\label{2p-pol}
\end{figure*}

The geometry considered for the results presented in this section is shown in
Figure~\ref{geom}. The plots of $Q/I$ vs. $U/I$ (polarization diagrams) are shown
for the line center wavelengths of the Li D lines in Figure~\ref{2p-pol}.
For the geometry considered, only the level-crossings
with $|\Delta\mu|=2$ are effective.
Therefore, in the following, we will only see the effects due to the first level-crossing
at 3.238 kG.

At the $^7$D$_2$ line center ($\lambda=6707.74416$\,\AA) we see a decrease in $Q/I$
up to a few hundred gauss (Hanle regime), with an initial increase followed by a decrease
in $U/I$ (see Figure~\ref{2p-pol}a). This is due to the Hanle effect
which operates in the line core. As we approach the level-crossing field strength
($B=3.238$ kG), we see an increase in the value of $Q/I$ and formation of a loop.
Indeed the level-crossing occurs over a narrow range of field strengths around 3.238 kG.
Within this narrow range if the field strength is varied by organizing a fine grid,
we get a strong variation in $Q/I$ and $U/I$, producing the polarization
diagram shown in Figure~\ref{2p-pol}a. This behavior is generic to all the polarization
diagrams shown in Figure~\ref{2p-pol}. Further discussion on the formation of loops
around the level-crossing field strengths can be found in LL04. Basically at the
level-crossing field strengths, the coherence between the overlapping magnetic
substates increases, resulting in the scattered $Q/I$
tending towards the non-magnetic value.
For kG fields, $U/I$ becomes zero because of the geometry considered. For fields stronger
than 10 kG (see Figure~\ref{2p-pol}b), $Q/I$ arises due
to Rayleigh scattering in strong magnetic fields, as discussed
by \citet[][Section 6, Figure 14]{bom97b}. The author states that in this case, for the
geometry considered (magnetic field along the line of sight), and for a
90\textdegree\ scattering, only the $\sigma$ components are scattered, if the atomic system
considered is a normal Zeeman triplet ($J=0\rightarrow1\rightarrow0$). Incidentally
we notice the same behavior in the case of $L=0\rightarrow1\rightarrow0$
transition (which is not a normal Zeeman triplet). It is interesting to note that
the $\pi$ components are not scattered in this case also.
The $Q/I$ changes sign and increases for fields up to 100 kG.
At the $^6$D$_2$ line center ($\lambda=6707.90232$\,\AA; see Figure~\ref{2p-pol}c)
the $Q/I$ for zero magnetic field case is approximately ten times smaller than the
$Q/I$ at the $^7$D$_2$ line center. This is due to the
relatively small abundance of $^6$Li. Due to an increased separation between
the magnetic components with the field strength, the $Q/I$ value
decreases. As in the case of $^7$D$_2$ we notice the formation of a loop near the
level-crossing field strength. When the field strength is increased beyond 10 kG,
we again notice Rayleigh scattering in strong magnetic fields (not shown in the figure).

The $^6$D$_1$ line ($\lambda=6708.05534$\,\AA; see Figure~\ref{2p-pol}d) is
intrinsically unpolarizable as it has $W_2=0$. Therefore the polarization remains
zero until the level-crossing field strength ($B=3.238$ kG) is reached. A further
increase in the field strength results in the formation of a loop and Rayleigh scattering
as already described for the $^7$D$_2$ and $^6$D$_2$ line center positions.

\section{Conclusions}
\label{conclu}
The theory of Hanle effect in a two-term atom was developed by LL04 
assuming a flat-spectrum for the incident unpolarized radiation using the density
matrix formalism. Though this theory is applicable to the entire range of magnetic
fields, it does not take into account the effects of PRD. \citet{smi11a} included
the effects of PRD by taking the redistribution matrix approach but their treatment
was limited to the linear Zeeman regime.
In the present paper we have generalized the
approach of \citet{smi11a} for magnetic fields of arbitrary strengths to include the 
Paschen-Back regime. However, our treatment ignores the effects of collisions.
Further the lower term is assumed to be unpolarized. The frequency dependence of
the incident radiation field is taken into account in our theory which is essentially
a relaxation of the flat-spectrum approximation of LL04. This enables us to properly
calculate the scattered Stokes profile shapes which was otherwise not possible with
the theory presented in LL04.

An example where the present theory has observable effects on the Sun is for the
lithium 6708\,\AA\ doublet. Since the fine structure splitting in this line system
is small (0.15\,\AA), Paschen-Back effects in scattering polarization become prominent
for magnetic fields that occur on the Sun. We have therefore tested our theory by
applying it to this lithium doublet and demonstrated the correctness of the formalism
by reproducing available benchmarks. We have explored the properties of the
redistribution matrix for the single scattering case, and clarified the effects of
Rayleigh scattering in strong fields when the magnetic field is along the line of sight.
This has given us an overview of the behavior of the polarization effects that can
occur as a result of PRD in the Paschen-Back regime.

We acknowledge the use of HYDRA cluster at the Indian Institute of Astrophysics for
computations in this work.

\label{lastpage}
\end{document}